# Emotion Recognition based on Third-Order Circular Suprasegmental Hidden Markov Model

Ismail Shahin
Department of Electrical and Computer Engineering
University of Sharjah
Sharjah, United Arab Emirates
E-mail: ismail@sharjah.ac.ae

*Abstract*—This work focuses on recognizing the unknown emotion based on the "Third-Order Circular Suprasegmental Hidden Markov Model (CSPHMM3)" as a classifier. Our work has been tested on "Emotional Prosody Speech and Transcripts (EPST)" database. The extracted features of EPST database are Mel-Frequency Cepstral Coefficients (MFCCs). Our results give average emotion recognition accuracy of 77.8% based on the CSPHMM3. The results of this work demonstrate that CSPHMM3 is superior to the Third-Order Hidden Markov Model (HMM3), Gaussian Mixture Model (GMM), Support Vector Machine (SVM), and Vector Quantization (VQ) by 6.0%, 4.9%, 3.5%, and 5.4%, respectively, for emotion recognition. The average emotion recognition accuracy achieved based on the CSPHMM3 is comparable to that found using "subjective assessment by human judges".

## I. INTRODUCTION

Emotion recognition is defined as recognizing the talking condition of a speaker speaking emotionally such as disgust, anger, and happiness. Emotion recognition has a large number of applications that emerge in "telecommunications, human robotic interfaces, smart call centers, and intelligent spoken tutoring systems". In "telecommunications, emotion recognition" can be seen in assessing the emotion of a caller for telephone reply facilities. Recognizing emotions can also be utilized in "human robotic interfaces" where robots can be trained to interact with humans and recognize human emotions. Further applications can be found in "smart call centers" where probable problems happening from unsatisfactory communications can be discovered by "emotion recognition systems". "Emotion recognition" can be utilized in "intelligent spoken tutoring systems" to feel and regulate to the emotions of students when students went into a tiresome state throughout tutoring meetings [1], [2], [3].

This work focuses on enhancing emotion recognition accuracy based on the "Third-Order Circular Suprasegmental Hidden Markov Model (CSPHMM3)" as a classifier. Our work has been tested on "Emotional Prosody Speech and Transcripts (EPST)" corpus.

The remaining of this paper is given as: Literature review is presented in Section II. The fundamentals of CSPHMM3 are given in Section III. The utilized database and extraction of features appear in Section IV. The algorithm of emotion recognition based on CSPHMM3 is explained in Section V. The achieved findings and the experiments are discussed in Section VI. Concluding remarks of our work are given in Section VII.

## II. LITERATURE REVIEW

Emotion recognition research branch is covered by a huge number of studies [4-12]. Morrison *et.al* [4] focused on enhancing the classification of emotion techniques using "ensemble or multi-classifier system (MCS)" methods. They also investigated the variances to recognize emotions of human beings that are taken from various techniques of acquisition. Casale *et.al* [5] introduced a new "feature vector" that assists in improving the grouping accuracy of "emotional/stressful" human conditions. The elements of such a "feature vector" are achieved from a "feature subset selection method based on genetic algorithm". Yogesh *et.al* [6] proposed "a new particle swarm optimization assisted biogeography-based algorithm" for feature selection. They performed their experiments utilizing "Berlin Emotional Speech database (BES), Surrey Audio-Visual Expressed Emotion database (SAVEE), and Speech Under Simulated and Actual Stress (SUSAS)" database.

Shahin focused in one of his previous research [7] on studying and enhancing "text-independent and speaker-independent talking condition identification in stressful and emotional environments" (entirely two independent environments) based on three independent and diverse classifiers: "Hidden Markov Models (HMMs), Second-Order Circular Hidden Markov Models (CHMM2s), and Suprasegmental Hidden Markov Models (SPHMMs)". His study demonstrated that "SPHMMs" lead "HMMs and CHMM2s" for emotion recognition in the two environments [7]. Shahin and Ba-Hutair [8] directed in one of their research on improving "text-independent and speaker-independent talking condition recognition in each of stressful and emotional environments" based on utilizing "Second-Order

Circular Suprasegmental Hidden Markov Models (CSPHMM2s)" as classifiers. Furthermore, one of the major goals of their work was to differentiate between "stressful talking environment and emotional talking environment based on CSPHMM2s". Their conclusion was that talking recognition in "stressful and emotional environments" based on "CSPHMM2s" is superior to that based on "HMMs, CHMM2s, and SPHMMs". In one of his preceding studies [9], Shahin operated emotions to identify the unknown speakers. He proposed a new framework to recognize speakers from their emotions based on HMMs. In another work by Shahin [10], he introduced, applied, and evaluated "speaker-dependent and text-dependent speaking style authentication (verification) systems" that admit or deny the identity claim of a speaking style based on SPHMMs. His results, based on SPHMMs, showed that the "average speaking style authentication performance is: 99%, 37%, 85%, 60%, 61%, 59%, 41%, 61%, and 57% corresponding, respectively, to the speaking styles: neutral, shouted, slow, loud, soft, fast, angry, happy, and fearful". To enhance speaker verification accuracy in "emotional talking environments", Shahin [11] proposed a two-stage approach that employs the emotion of speaker cues "(text-independent and emotion-dependent speaker verification problem) based on both HMMs and SPHMMs as classifiers". This framework is comprised of two cascaded stages that combine and integrate "emotion recognizer" followed by a "speaker recognizer" into one recognizer. His approach has been evaluated on two diverse and separate emotional speech databases: his collected database and "Emotional Prosody Speech and Transcripts (EPST)" database. The results of his work present that his introduced approach yields better results with a significant enhancement over prior studies and other approaches such as "emotion-independent speaker verification approach and emotion-dependent speaker verification approach based completely on HMMs". In a new work by Shahin and Bou Nassif [12], they aimed at enhancing emotion recognition accuracy based on "Third-Order Hidden Markov Models (HMM3s)" as a classifier. Their work has been assessed on EPST database. "The extracted features of EPST database are Mel-Frequency Cepstral Coefficients (MFCCs)". Their results gave an average emotion recognition accuracy of 71.8%. Their results indicated that HMM3s are superior to "First-Order Hidden Markov Models (HMM1s) and Second-Order Hidden Markov Models (HMM2s)" by 14.0% and 5.7%, respectively, for emotion recognition accuracy.

In this work, we aim at improving emotion recognition accuracy based on the "Third-Order Circular Suprasegmental Hidden Markov Model (CSPHMM3)" as a classifier. Our work has been evaluated on EPST corpus. The extracted features that have been used in this work are called "Mel-Frequency Cepstral Coefficients (MFCCs)".

### III. FUNDAMENTALS OF CSPHMM3

"Third-Order Circular Suprasegmental Hidden Markov Model has been developed from acoustic Third-Order Hidden Markov Model (HMM3)". Shahin [13] proposed, applied, and evaluated HMM3 to improve the dropped "text-independent speaker identification accuracy in a shouted talking environment".

#### A. Basics of HMM3

In "HMM1, the underlying state sequence is a first-order Markov chain" where the "stochastic process is specified by a 2-D matrix of a priori transition probabilities ($a_{ij}$)" between states $s_i$ and $s_j$ where $a_{ij}$ is given as [14],

$$\text{"}a_{ij} = \text{Prob}\left(q_t = s_j \middle| q_{t-1} = s_i\right)\text{"} \quad (1)$$

In "HMM2, the underlying state sequence is a second-order Markov chain" where the "stochastic process is described by a 3-D matrix ($a_{ijk}$)". Hence, the "transition probabilities in HMM2" are given as [15],

$$a_{ijk} = \text{Prob}\left(q_t = s_k \middle| q_{t-1} = s_j, q_{t-2} = s_i\right) \quad (2)$$

"with the constraints",

$$\sum_{k=1}^{N} a_{ijk} = 1 \qquad N \geq i, j \geq 1$$

In "HMM3, the underlying state sequence is a third-order Markov chain" where the "stochastic process is stated by a 4-D matrix ($a_{ijkw}$)". Subsequently, the "transition probabilities in HMM3" are given as [13],

$$a_{ijkw} = \text{Prob}\left(q_t = s_w \middle| q_{t-1} = s_k, q_{t-2} = s_j, q_{t-3} = s_i\right) \quad (3)$$

"with the constraints",

$$\sum_{w=1}^{N} a_{ijkw} = 1 \qquad N \geq i, j, k \geq 1$$

The probability of the state sequence, $Q \triangleq q_1, q_2, ..., q_T$, is expressed as:

$$\text{Prob}(Q) = \Psi_{q_1} a_{q_1 q_2 q_3} \prod_{t=4}^{T} a_{q_{t-3} q_{t-2} q_{t-1} q_t} \quad (4)$$

where "$\Psi_i$ is the probability of a state $s_i$ at time $t = 1$, $a_{ijk}$ is the probability of the transition from a state $s_i$ to a state $s_k$ at time $t = 3$". $a_{ijk}$ can be processed from equation (2). Thus, the initial parameters of HMM3 can be attained from the trained HMM2.

Given a sequence of observed vectors, $O \triangleq O_1, O_2, ..., O_T$, the joint state-output probability is expressed as [13]:

$$\text{Prob}(Q, O | \lambda) = \Psi_{q_1} b_{q_1}(O_1) a_{q_1 q_2 q_3} b_{q_3}(O_3) \prod_{t=4}^{T} a_{q_{t-3} q_{t-2} q_{t-1} q_t} b_{q_t}(O_t) \quad (5)$$

Readers can get more details about the three models from [13], [14], [15].

#### B. CSPHMM3

Within "Third-Order Circular Hidden Markov Model (CHMM3), prosodic and acoustic information can be merged into CSPHMM3" as given by the formula [16],

$$\log P\left(\lambda^v_{CHMM3s}, \Psi^v_{CSPHMM3s} \mid O\right) = (1-\alpha)\cdot \log P\left(\lambda^v_{CHMM3s} \mid O\right) \quad (6)$$
$$+ \alpha \cdot \log P\left(\Psi^v_{CSPHMM3s} \mid O\right)$$

where "$\lambda^v_{CHMM3}$ is the acoustic third-order circular hidden Markov model of the $v^{th}$ emotion and $\Psi^v_{CSPHMM3}$ is the suprasegmental third-order circular hidden Markov model of the $v^{th}$ emotion". Figure 1 displays an example of a basic structure of CSPHMM3 that has been formed from CHMM3. This figure is comprised of "six third-order acoustic hidden Markov states: $q_1, q_2,…, q_6$ positioned in a circular form. $p_1$ is a third-order suprasegmental state that is made up of $q_1$, $q_2$, and $q_3$. $p_2$ is a third-order suprasegmental state which is composed of $q_4$, $q_5$, and $q_6$". "The suprasegmental states $p_1$ and $p_2$ are located in a circular form. $p_3$ is a third-order suprasegmental state that is comprised of $p_1$ and $p_2$".

## IV. Speech Dataset and Extraction of Features

### A. Speech Dataset

In this research, our work has been tested on a worldwide speech dataset termed "Emotional Prosody Speech and Transcripts (EPST)". EPST contains 8 trained speakers ("3 actors and 5 actresses") who utter a sequence of "semantically neutral utterances comprising of dates and numbers" uttered in 15 diverse emotions. These emotions are "neutral, hot angry, cold angry, panicky, anxious, despaired, sad, elated, happy, interested, bored, shameful, proud, disgusted, and contempt" [17]. In our work, only 20 distinct utterances "(10 utterances were utilized for training and the rest were utilized for testing) spoken by 8 speakers (5 speakers were utilized for training and the rest were utilized for testing) talking in 6 different emotions were used. The emotions are neutral, hot angry, sad, happy, disgusted, and panicky".

### B. Extraction of Features

In the present work, the "phonetic content of speech signals" in EPST database is characterized by "Mel-Frequency Cepstral Coefficients (static MFCCs) and delta Mel-Frequency Cepstral Coefficients (delta MFCCs)". Such coefficients have been mainly used in many studies in the fields of "speech recognition [18], [19], speaker recognition [20], [21], and emotion recognition" [12], [22]. In this work, "MFCC feature analysis" is utilized to establish the "observation vectors in each of HMM3 and CSPHMM3". The computation of MFCC is shown in the block diagram of Fig. 2 [23].

A 32-dimension "MFCC (16 static MFCCs and 16 delta MFCCs) feature analysis" is used to produce the "observation vectors" in every model of "HMM3 and CSPHMM3". The "number of conventional states, $N$, in every model is 6 and the number of suprasegmental states is two (each suprasegmental state is made up of three conventional states) in CSPHMM3 with a continuous mixture observation density has been chosen for each model".

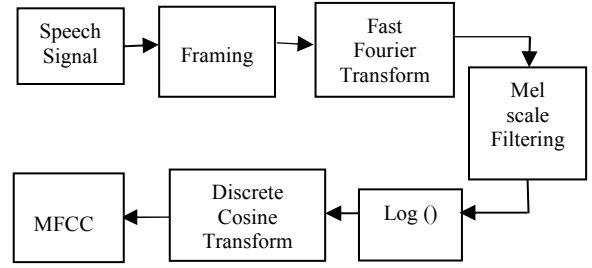

Fig. 2. Block diagram of MFCC algorithm

## V. Algorithm of Emotion Recognition Based on CSPHMM3

The "training phase" of CSPHMM3 is very identical to the training phase of the conventional CHMM3. In the training phase of "CSPHMM3, suprasegmental third-order circular model is trained on top of acoustic third-order circular model". In the "training phase of CHMM3", one reference model per emotion has been built utilizing "5 of the 8 speakers speaking 10 utterances with a replication of 2 times per utterance". The overall number of utterances that has been utilized in this phase to construct the models of the six emotions (each emotion is represented by one model) is "600 (5 speakers × 10 utterances × 2 replicates/utterance × 6 emotions)".

In the "test phase" of CSPHMM3, every one of the "3 remaining speakers uses different 10 utterances with a replication of 2 times per utterance under each emotion (text-independent and speaker-independent experiments). The total number of utterances that has been utilized in this phase is 360 (3 speakers × 10 utterances × 2 times/utterance × 6 emotions)". The probability of generating each utterance is calculated as,

$$E^* = \arg\max_{6 \geq e \geq 1}\left\{P\left(O \mid \lambda^e_{CHMM3s}, \Psi^e_{CSPHMM3s}\right)\right\} \quad (7)$$

where "$E^*$ is the index of the recognized emotion, $O$ is the observation vector that belongs to the unknown emotion, ($\lambda^e$) is the $e^{th}$ CHMM3 emotion model, and ($\Psi^e$) is the $e^{th}$ CSPHMM3 emotion model".

## VI. Results and Discussion

In this work, emotion recognition has been evaluated based on "CHMM3 and CSPHMM3" using a renowned speech database named "Emotional Prosody Speech and Transcripts".

Table I illustrates emotion recognition accuracy using EPST based on "CHMM3 and CSPHMM3". This table yields average emotion recognition accuracy of 73.4% and 77.8% based on "CHMM3 and CSPHMM3", respectively. It is clear that average emotion recognition accuracy based on CSPHMM3 is greater than that based on CHMM3 by 6.0%. It is evident from this table that the "suprasegmental model CSPHMM3 is superior to its corresponding acoustic model CHMM3" for emotion recognition.

A "statistical significance test" has been implemented to investigate whether emotion recognition accuracy difference (emotion recognition accuracy based on CSPHMM3 and that based on CHMM3) is real or only comes from statistical variations. The "statistical significance test" has been used based on the "Student's *t* distribution test" as given by,

$$t_{\text{model } x, \text{model } y} = \frac{\overline{X}_{\text{model } x} - \overline{X}_{\text{model } y}}{SD_{\text{pooled}}} \quad (8)$$

where "$\overline{X}_{\text{model } x}$ is the mean of the first sample (model *x*) of size *n*, $\overline{X}_{\text{model } y}$ is the mean of the second sample (model *y*) of the same size, and $SD_{\text{pooled}}$ is the pooled standard deviation of the two samples (models *x* and *y*)" given as,

$$SD_{\text{pooled}} = \sqrt{\frac{SD^2_{\text{model } x} + SD^2_{\text{model } y}}{2}} \quad (9)$$

where "$SD_{\text{model } x}$: is an estimate of the standard deviation of the average of the first sample (model *x*) of size *n* and $SD_{\text{model } y}$ is an estimate of the standard deviation of the average of the second sample (model *y*) of the same size".

The "calculated *t* value" between CSPHMM3 and CHMM3 is computed based on Table I. The computed value is $t_{\text{CSPHMM3, CHMM3}} = 1.864$ which is greater than the "tabulated critical value $t_{0.05} = 1.645$ at 0.05 significant level". Thus, it is apparent that CSPHMM3 outperforms CHMM3 for emotion recognition.

Table II exhibits a "confusion matrix" that characterizes the "percentage of confusion" of a test emotion with the other emotions based on CSPHMM3. This table states that:

1. The most easily recognizable emotion is neutral (96.5%). Consequently, the highest emotion recognition accuracy is neutral.

2. The least easily recognizable emotion is hot anger (64.5%). Therefore, the least emotion recognition accuracy is hot anger.

3. The last column "Panic", for instance, explains that 5% of the utterances that were portrayed in a panic emotion were assessed as spoken in a hot anger emotion, 3% of the utterances that were produced in a panic emotion were recognized as generated in a happy emotion. This column shows that panic emotion has the maximum "confusion percentage" with sad emotion (9%). Hence, panic emotion is largely confused with sad emotion. This column also displays that panic emotion has the least "confusion percentage" with neutral emotion (0%). So, panic emotion is absolutely not confused at all with neutral emotion. This column indicates that 75.5% of the utterances that were produced in a panic emotion were recognized properly.

Emotion recognition accuracy based on "CSPHMM3 has been contrasted with that based on the state-of-the-art classifiers and models such as Gaussian Mixture Model (GMM) [24], Support Vector Machine (SVM) [25], and Vector Quantization (VQ)" [26]. Emotion recognition accuracy based on GMM, SVM, and VQ is 74.2%, 75.2%, and 73.8%, respectively. It is apparent from this experiment that CSPHMM3 leads GMM, SVM, and VQ by 4.9%, 3.5%, and 5.4%, respectively, for emotion recognition.

An "informal subjective assessment" for emotion recognition using EPST dataset has been conducted using 10 human non-professional adult listeners. In this assessment, a sum of 480 utterances (8 speakers × 6 emotions × 10 utterances) have been utilized. These listeners are enquired to recognize the unknown emotion. The average emotion recognition accuracy using EPST database is 71.4%. This average is close to the obtained average based on CSPHMM3 (77.8%).

## VII. CONCLUDING REMARKS

In this work, we utilize CHMM3 and CSPHMM3 as classifiers to identify the unknown emotion using a very famous speech dataset called Emotional Prosody Speech and Transcripts. Some concluding remarks can be drawn in this work. Firstly, "CSPHMM3" is superior to each of "CHMM3, GMM, SVM, and VQ" for emotion recognition. Secondly, the maximum emotion recognition accuracy happens when speakers speak neutrally. Finally, the minimum emotion recognition accuracy takes place when speakers speak angrily.

There are some limitations in this work. First, EPST dataset has limited number of speakers. Second, the achieved emotion recognition accuracy based on CSPHMM3 is imperfect. Our plan for future work is to use up-to-date classifiers to improve emotion recognition accuracy.

## ACKNOWLEDGMENT

The author of this work wishes to express his thanks and gratitude to the "University of Sharjah" for supporting this work through the "two competitive research projects" entitled "Emotion Recognition in each of Stressful and Emotional Talking Environments Using Artificial Models, No. 1602040348-P" and "Capturing, Studying, and Analyzing Arabic Emirati-Accented Speech Database in Stressful and Emotional Talking Environments for Different Applications, No. 1602040349-P".

Table I. Emotion Recognition Accuracy using EPST based on "CHMM3 and CSPHMM3"

| Model | Gender | "Emotion recognition accuracy under each emotion" (%) | | | | | |
|---|---|---|---|---|---|---|---|
| | | "Neutral" | "Hot Anger" | "Sadness" | "Happiness" | "Disgust" | "Panic" |
| CHMM3 | Male | 96 | 59 | 72 | 68 | 72 | 72 |
| | Female | 96 | 58 | 72 | 69 | 73 | 74 |
| | Average | 96.0 | 58.5 | 72 | 68.5 | 72.5 | 73.0 |
| CSPHMM3 | Male | 97 | 65 | 76 | 77 | 77 | 75 |
| | Female | 96 | 64 | 78 | 77 | 76 | 76 |
| | Average | 96.5 | 64.5 | 77.0 | 77.0 | 76.5 | 75.5 |

Table II. Confusion Matrix of Emotion Recognition based on CSPHMM3

| Talking condition | "Percentage of confusion of a test emotion with the other emotions" (%) | | | | | |
|---|---|---|---|---|---|---|
| | "Neutral" | "Hot Anger" | "Sadness" | "Happiness" | "Disgust" | "Panic" |
| Neutral | **96.5** | 3 | 3 | 6 | 2.5 | 0 |
| Hot Anger | 0 | **64.5** | 5 | 2 | 8 | 5 |
| Sadness | 0 | 10 | **77** | 2 | 3 | 9 |
| Happiness | 1.5 | 2 | 2 | **77** | 2 | 3 |
| Disgust | 1 | 6.5 | 5 | 6 | **76.5** | 7.5 |
| Panic | 1 | 14 | 8 | 7 | 8 | **75.5** |

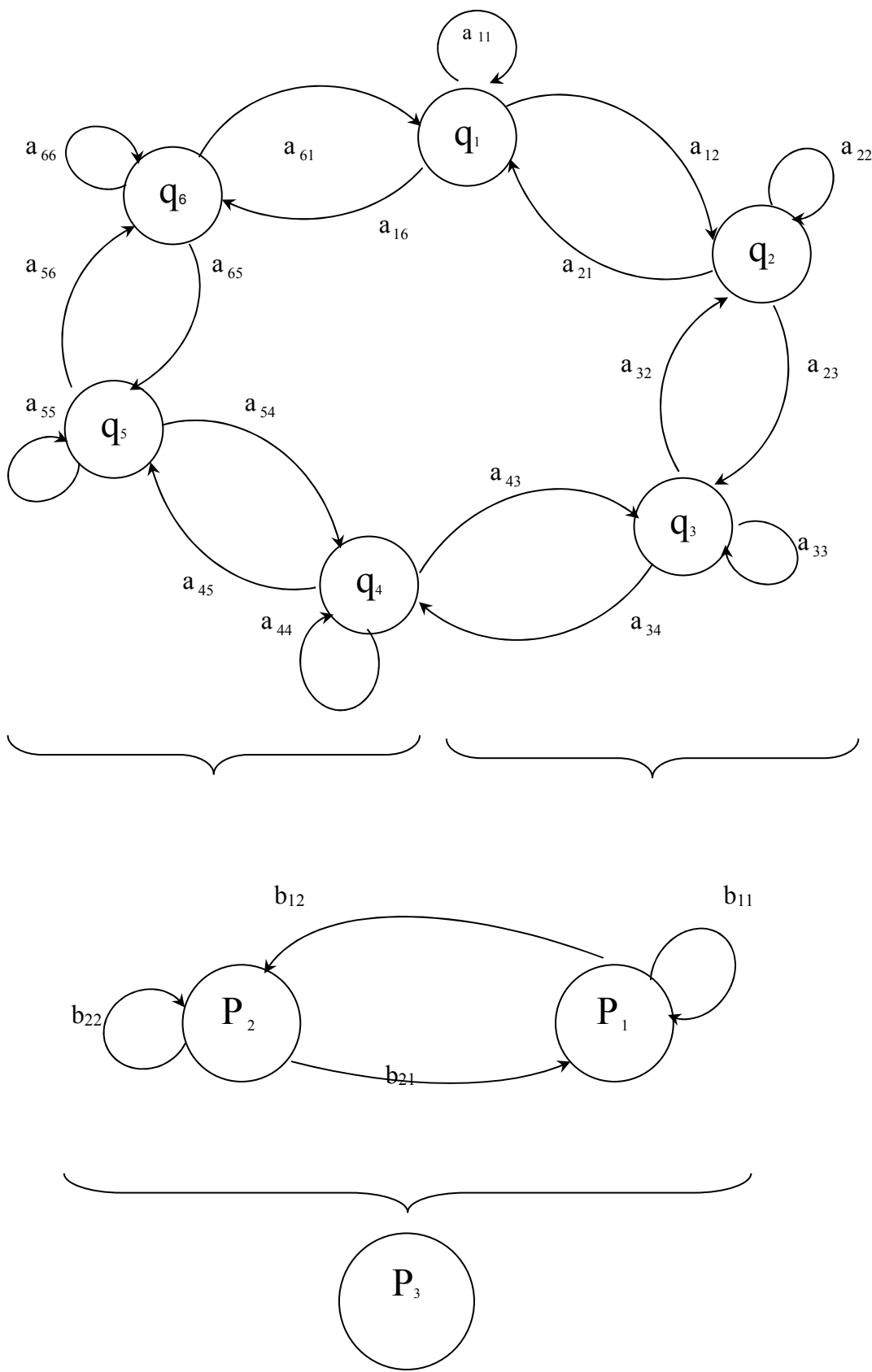

Fig. 1. Basic structure of CSPHMM3 derived from CHMM3